\documentclass[aps,reprint,superscriptaddress,amsmath,amssymb,prl,showpacs]{revtex4-1} 

\usepackage{graphicx}
\usepackage{dcolumn}
\usepackage{bm}
\usepackage{bm}
\usepackage{amssymb,amsfonts,amsmath}
\usepackage[normalem]{ulem}
\usepackage{epstopdf}

\newcommand{\bra}[1]{\left(#1\right)}

\begin{document}


\title{Theory of phase separation and polarization for dissociated ionic liquids}

\author{Nir Gavish}
\email{ngavish@tx.technion.ac.il}
\affiliation{Department of Mathematics, Technion--Israel Institute of Technology, Technion City 32000, Israel}

\author{Arik Yochelis}\email{yochelis@bgu.ac.il}
\affiliation{Department of Solar Energy and Environmental Physics, Swiss Institute for Dryland Environmental and Energy Research, Blaustein Institutes for Desert Research (BIDR), Ben-Gurion University of the Negev, Sede Boqer Campus, Midreshet Ben-Gurion 84990, Israel}

\date{\today}

\begin{abstract}
Room temperature ionic liquids are attractive to numerous applications and particularly, to renewable energy devices. As solvent free electrolytes, they demonstrate a paramount connection between the material morphology and Coulombic interactions: unlike dilute electrolytes, the electrode/RTIL interface is a product of both electrode polarization and spatiotemporal bulk properties. Yet, theoretical studies have dealt almost exclusively with independent models of morphology and electrokinetics. In this work, we develop a novel Cahn-Hilliard-Poisson type mean-field framework that couples morphological evolution with electrokinetic phenomena. Linear analysis of the model shows that spatially periodic patterns form via a finite wavenumber instability, a property that cannot arise in the currently used Fermi-Poisson-Nernst-Planck equations. Numerical simulations in above one-space dimension, demonstrate that while labyrinthine type patterns develop in the bulk, stripe patterns emerge near charged surfaces. The results qualitatively agree with empirical observations and thus, provide a physically consistent methodology to incorporate phase separation properties into an electrochemical framework.
\end{abstract}
                              
\maketitle

{\em Introduction--}
High energy consumption stresses the need for the development of renewable and clean devices that will ease the transition from reliance on fossil fuels towards independent alternatives. Room temperature ionic liquids (RTILs) are attractive to several technological applications and in particular to renewable energy devices~\cite{SilComp:2006,Wishart2009956,B817899M,Brandt2013554,Li201483}, due to their high charge density and tunable anion/cation design, low vapor pressure, and wide electrochemical windows. RTILs are molten salts and thus propose superior properties to electrolyte solutions when incorporated into devices such as batteries, supercapacitors, dye-sensitized solar cells. Even though RTILs resemble highly concentrated electrolytes, they also exhibit fundamental physicochemical differences: (i) Electrical double layer (EDL) structure is observed to be profoundly different from that in traditional (dilute) electrolytes~\cite{Perkin20125052,Mezger2008424,Atkin20075162,Fukushima20032072,Carmichael2001795}; (ii) RTIL bulk often exhibit a nanostructure which displays laminar, bicontinuous, and sponge-like morphologies~\cite{Triolo20074641,Xiao200813316,Li20084375,Atkin20084164,Busico19821033,Abdallah20003053,Triolo20074641,Atkin20084164}.

Although the nanostructuring of the RTIL is fundamental for device efficiency, it is poorly understood at both mechanistic and engineering levels~\cite{Atkin20084164,Hayes20101709}. Traditionally models of either atomistic or continuum type, consider separately EDL structuring by electrode polarization and bulk morphology~\cite{fedorov2014ionic,hayes2015structure}.  Empirical and theoretical evidences, however, strongly suggest that a combined theory should be advanced, cf.~\cite{Yochelis2015} and the references therein. Atomistic (force field) methods such as molecular dynamics, allow access to relatively realistic properties of RTILs. These methods,  however, are limited to relatively small systems due to a finite number of molecules that can be traced simultaneously~\cite{Merlet201315781,Nagasawa201413008}. On the other hand, although mean--field formulations do not offer atomistic insights they are amenable to extended analytical and numerical computations and thus may offer fundamental understanding of the system behaviour at larger scales.

Respectively, modeling of bulk structure was advanced by formulating the problem via the gradient flow (e.g., Flory-Huggins) approach~\cite{Aerov201015066,Aerov2012}.  These models predict amphiphilic-type bulk structures, in agreement with empirical data~\cite{Triolo20074641,Xiao200813316,Li20084375,Atkin20084164}. Nevertheless, these models do not account for an external electrical field and thus do not provide insights to the EDL structure. On the other hand, charge layering of the EDL was captured via incorporation of the over--screening effect~\cite{Bazant2011,Lee2015} through the so called Fermi-Poisson-Nernst-Planck (FPNP) framework~\cite{Santangelo:2006}, but while assuming a structureless bulk in the absence of the applied potential. These equations allowed insights into the ion-pairing evolutions~\cite{Gebbie2013,Lee2015159} and moreover, stressed the need in connecting the EDL structure with the bulk nature~\cite{Yochelis2014,Yochelis2014b}.  
 
In this letter, we develop a mean-field theory that combines phase separation and ionic transport via an Onsager framework~\cite{hilliardspinodal,elliott1997diffusional}, while incorporating finite size effects and Coulombic interactions. The resulting model  bears similarity to the Ohta-Kawasaki formulation for morphology development driven by nonlocal interactions~\cite{ohta1986equilibrium,kawasaki1988equilibrium,choksi2005diblock}, and in fact offers an alternative for the FPNP approach~\cite{Santangelo:2006,Bazant2011,Lee2015}. Using linear stability analysis, we show that the bulk morphology emerges via a finite wavenumber instability type~\cite{Cross1993851}. Below the onset and upon applied potential, we observe also the crowding effect, and therefore our approach is a consistent alternative to the FPNP model. Moreover, numerical simulations show that the same phenomenology  persists in higher space dimensions. Finally, we address implications to empirical observations and outlook other electrochemical systems.

{\em Coupling phase separation and polarization --} RTILs are fundamentally distinct from dilute electrolytes, which obey the Poisson--Nernst-Planck description, due to the absence of a solvent, such as water. Consequently, we start by considering a system of a symmetric RTIL of monovalent anions ($n$) and cations ($p$), confined in between two flat parallel electrodes. The salt ions are assumed be fully dissociated and thus, their local volume fractions, $0\le (p,n) \le 1$, preserve the uniform density constraint:
\begin{equation}\label{eq:uniform_density}
n({\bf x})+p({\bf x})=1,
\end{equation}
where~$\bf x \in \mathbb{R}^3$. In the absence of a solvent (namely, large concentration gradients) and since the molecules are charged, the driving forces for ionic transport and structural evolution of the RTIL are attributed to short-range (phase separating) interactions and to long-range Coulombic interactions. Thus, the mean-field free energy of the system is given by
\begin{subequations} \label{eq:energy_total}
	\begin{equation}\label{eq:energy}
	E=E_{\rm CH}+E_{\rm C},
	\end{equation}
where
\begin{eqnarray}\label{eq:electrostatic+entropy}
\label{eq:entropy}
E_{\rm CH}&=&\bar{c}\int  f_m(p,n)+ \frac{E_0\kappa^2}4 \left(\left|\nabla p\right|^2+\left|\nabla n\right|^2\right)\,d{\bf x},\\
E_{\rm C}&=&\int q\bar{c}(p-n)\phi-\frac12\epsilon|\nabla \phi|^2\,d{\bf x},
\label{eq:electrostatic}
\end{eqnarray}
and
\[
f_m(p,n)=k_BT \left[p\ln \frac p2 +n\ln \frac n2\right]+\beta np.
\]
\end{subequations}
Here,~$\bar{c}$ is the concentration of the anions and of the cations in the mixture,~$k_B$ is the Boltzmann constant,~$T$ is the temperature,~$\beta$ is interaction parameter for the anion/cation mixture and has units of energy,~$E_0\kappa^2/4$ is the gradient energy coefficient where~$E_0$ has units of energy and~$\kappa$ has units of length. 

In~(\ref{eq:energy}), the first term $E_{\rm CH}$ stands for the Cahn-Hilliard energy and accounts for the energetic cost of anion/cation mixing and for composition inhomogeneities~\cite{huggins1942theory,Flory_book,cahn1958free}. When~$\beta>\beta_c=2k_BT$, the function $f_m(p,n\equiv 1-p)$ takes the form of a double well potential, thus driving phase separation.
The second component,~$E_{\rm C}$ of~\eqref{eq:energy}, stands for the electrostatic energy where $\phi$ is the electric potential,~$q$ is the elementary charge, and~$\epsilon$ is the permittivity. Requiring that~$\phi$ is a critical point of the action yields Poisson's equation,
\begin{equation}\label{eq:Poisson}
\dfrac{\delta E}{\delta \phi}=q\bar{c}(p-n)+\epsilon \nabla^2\phi=0.
\end{equation}
Notably, Eqs.~\eqref{eq:entropy} and~\eqref{eq:Poisson}, in fact comprise the Ohta-Kawasaki model, a nonlocal Cahn-Hilliard model for the structure of diblock copolymers mixtures~\cite{ohta1986equilibrium,kawasaki1988equilibrium,choksi2005diblock}.


{\em Ion transport --}
The uniform density constraint~\eqref{eq:uniform_density} for RTILs, implies that ion transport is governed by an inter-diffusion process and not by a standard diffusion process, i.e., not by a random walk of isolated ions in a solvent. Following Onsager's framework~\cite{hilliardspinodal}, the equations of motion for cations and anions read:
\begin{subequations}\label{eq:motion_gen}
\begin{equation}
\partial_t\left[\begin{array}{c} p\\n\end{array}
\right]=\nabla\cdot \left({\bf L}({\bf x})\,\left[\nabla \dfrac{\delta E}{\delta p},\nabla \dfrac{\delta E}{\delta n}\right]^T
\right),
\end{equation}
where~${\bf L}({\bf x})$ is the $2\times2$ matrix of coefficients and the superscript $T$ stands for transpose.
Onsager's reciprocal relations combined with~\eqref{eq:uniform_density}, leads to the choice of the Onsager matrix~\cite{elliott1997diffusional} 
\begin{equation}
{\bf L}({\bf x})=\frac{M\,p({\bf x})n({\bf x})}{\bar{c}} \left[\begin{array}{rr}1&-1\\-1& 1\end{array}\right],
\end{equation}
where $M$ is the mobility coefficient~\cite{de1980dynamics,brochard1983polymer}.
\end{subequations}
Introduction of the standard dimensionless variables 
\[
\tilde\phi=\frac{q}{k_BT}\phi,~\tilde {\bf x}=\frac{\bf x}{\lambda},~\lambda=\sqrt{\frac{\epsilon k_BT}{2q^2\bar{c}}},~\tilde t=\frac{t}{\tau},~\tau=\frac{\lambda^2}{M k_BT},
\]
leads, after omitting the tildes, to the dimensionless form of~\eqref{eq:Poisson} and~\eqref{eq:motion_gen}: 
\begin{subequations}\label{eq:rtil_model}
\begin{equation}
\label{eq:rtil_model_phi}
\nabla^2 \phi= 1-2p,
\end{equation}
and
\begin{eqnarray}
\label{eq:rtil_model_p}
\partial_t p&=& -\nabla\cdot \mathbf{J}, \qquad \partial_t n=-\partial_t p, \\
\nonumber \mathbf{J}&=&\left(p-1\right) p\nabla \left(-\sigma\nabla^2 p+\log \frac{p}{1- p}+\chi(1-2 p)+2\phi\right).
\end{eqnarray}
\end{subequations}
Here, $\lambda$ is the Debye length, $\sigma=\frac{E_0}{k_BT}\frac{\kappa^2}{\lambda^2}$ controls the competition between short- and long-range interactions, and~$\chi=\beta/(k_BT)$ is the Flory parameter.
Eqs.~\eqref{eq:rtil_model} are supplemented with boundary conditions of fixed potential at the electrodes and Neumann (i.e., no-flux) for $p$ and $n$:  
\begin{equation}\label{eq:BC}
\phi(x=0)=-V/2,~~ \phi(x=d)=V/2,~~\mathrm{and}~~ \mathbf{J}\lvert_{\partial\Omega}=0,
\end{equation}
where $\Omega$ is the volume boundary and $d$ is the distance between electrodes; at the boundaries of the rest of the volume (that is in the $y-z$ planes), we take a Neumann boundary condition, i.e., $\nabla\phi|_{\partial\Omega y-z}=0$.

\begin{figure}[bp]
	\includegraphics[width=3in]{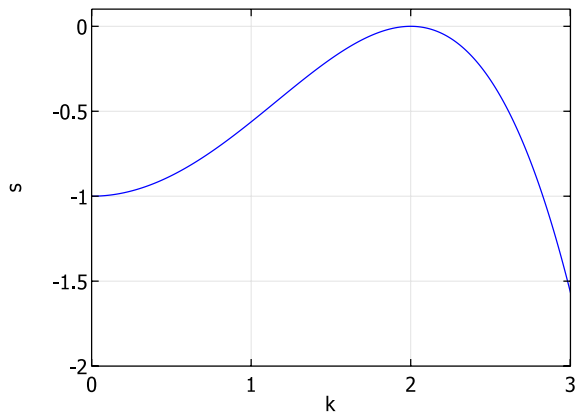}
	\caption{Dispersion relation (\ref{eq:disp}) showing the finite wavenumber instability onset at $\sigma=\sigma_c= 1/4$ and $\chi=3$.}
	\label{fig:RTILdisp}
\end{figure}

{\em Stability and numerical analysis--}
For the sake of analysis, we consider first an infinite one-space dimensional domain for which the electrode polarization effects vanish. Under such conditions, Eq.~\ref{eq:rtil_model_p} reduces to
%
\begin{equation}\label{eq:rtil_model_p_1D}
\partial_tp=\partial_x\left[\left(1-p\right) p\left(-\sigma\partial_x^3 p-2\chi\partial_x p+2\partial_x\phi\right)\right]+\partial_x^2p.
\end{equation}
Next, we substitute the expansions
\begin{equation}
p=  p_0 +\sum_{i=1}^\infty \varepsilon^i p_i,\quad \phi=  \sum_{i=1}^\infty \varepsilon^i \phi_i,
\end{equation}
into (\ref{eq:rtil_model_p_1D}) and collect terms up to the first order in  $\varepsilon \ll 1$. Noting that terms of the type $(\partial_x p)(\partial_x \phi)\sim o(\varepsilon^2)$, we identify using
\begin{equation}\label{eq:pert}
p_1 = e^{st+ikx} + c.c.,
\end{equation}
the stability properties of the equilibrium uniform state $(p_0,\phi_0)=(1/2,0)$; $s$ is the temporal growth rate of periodic perturbations associated with wavenumbers $k$~\cite{Cross1993851}. The resulting dispersion relation reads: 
\begin{equation}\label{eq:disp}
s=- 1+\bra{\dfrac{\chi}{2} - 1} k^2 - \dfrac{\sigma}{4} k^4.
\end{equation}
The uniform state is stable if for all~$k$, $s<0$, while the instability corresponds to a band of wavenumbers for which growth rate becomes $s>0$; the dispersion relation here is always real. The instability onset is obtained by seeking for a critical wavenumber for which $s(k=k_c)=0$, $s(k \neq k_c)<0$ and $\mathrm{d}s/\mathrm{d}k=0$. Accounting these conditions, we find that for $\chi > 2$ and $\sigma=\sigma_c$ the instability is of finite wavenumber type with
\begin{equation}\label{eq:onst}
\sigma_c = \dfrac{\bra{\chi-2}^2}{4},\quad \mathrm{and} \quad k_c=\dfrac{2}{\sqrt{\chi-2}},
\end{equation}
as demonstrated in Fig.~\ref{fig:RTILdisp}. Finite wavenumber instability is characteristic of systems with auto-catalytic kinetics, such as Turing's activator-inhibitor scenario, and distinct from the long wavenumebr instability of Cahn--Hilliard equations that describe coarsening (phase separation) evolution, i.e., a situation for which a typical spatial periodicity is not preserved, for details see~\cite{Cross1993851}.
\begin{figure}[tp]
\large	(a)\includegraphics[width=3in]{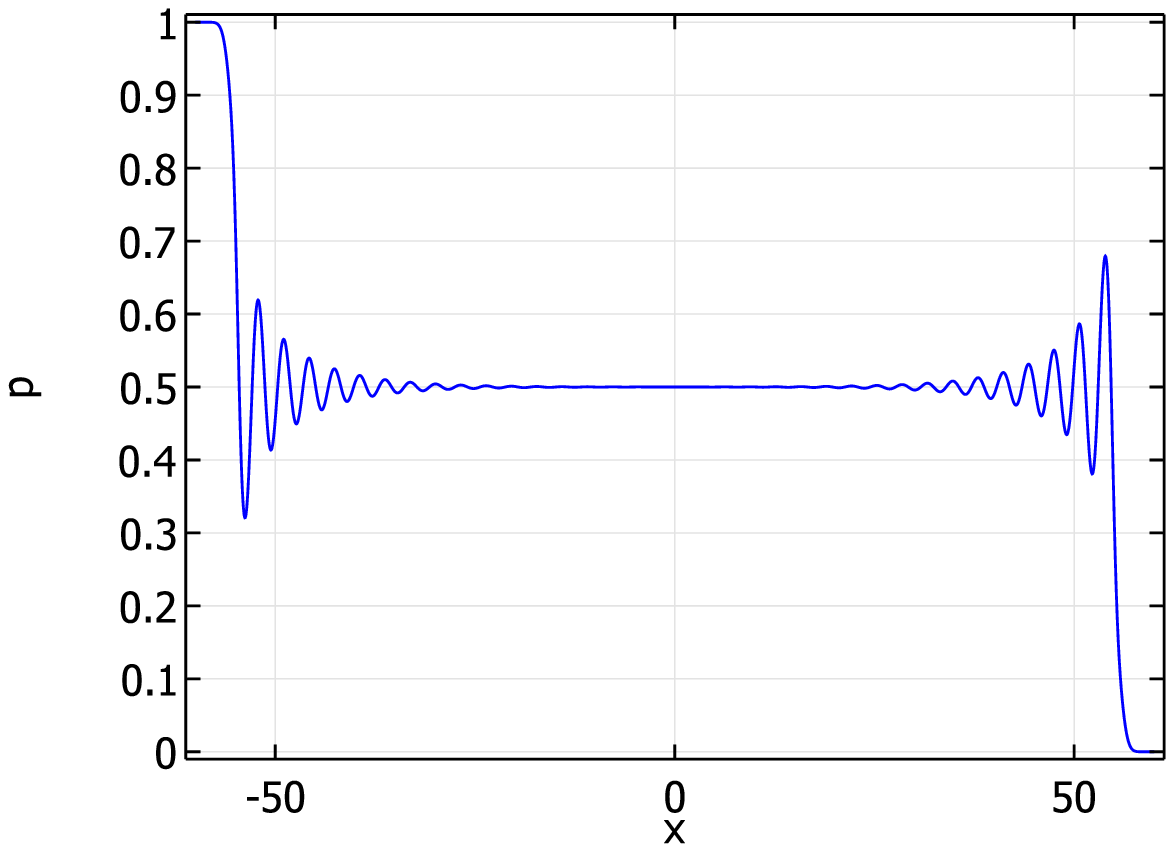}
	(b)\includegraphics[width=3in]{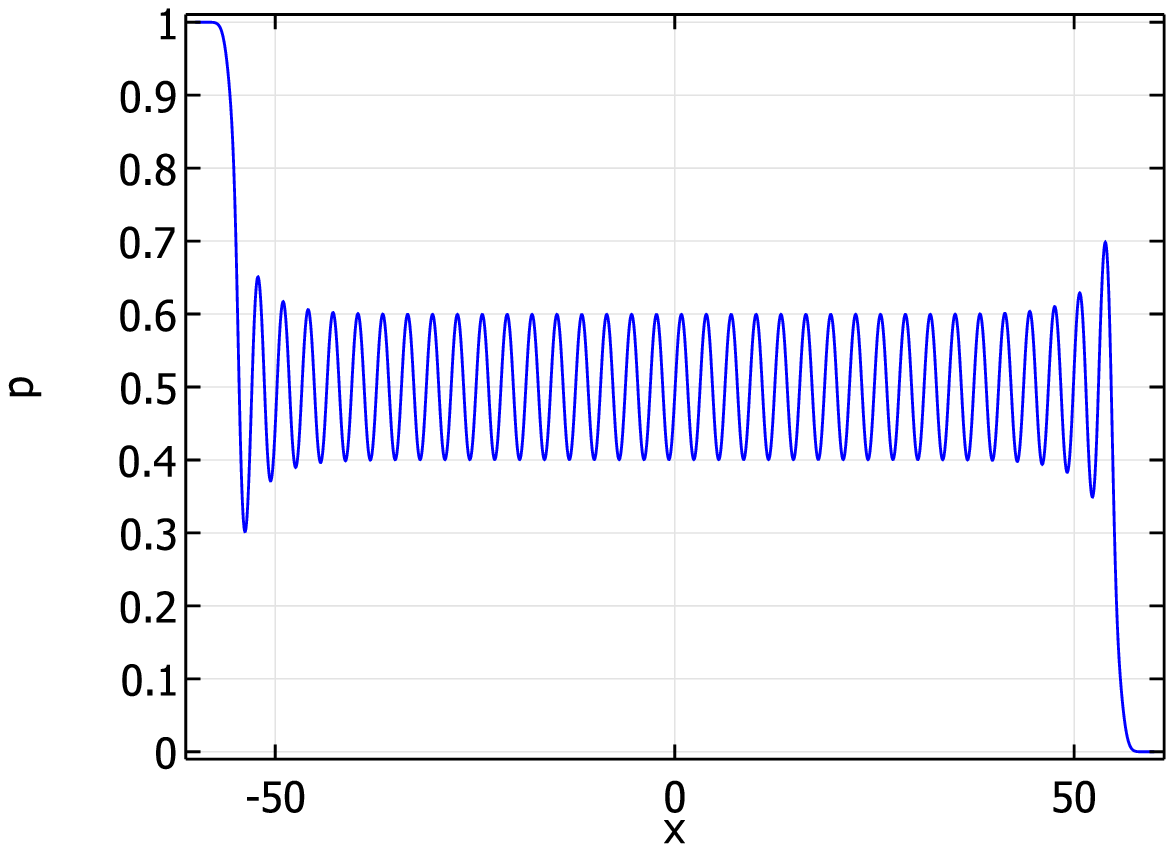}
	\caption{Numerical solutions of model equations~(\ref{eq:rtil_model}) in 1D, below (a) and above (b) the instability onset, showing the asymptotic states for $\chi=3$ and (a) $\sigma=0.255>\sigma_c= 1/4$, (b) $\sigma=0.24<\sigma_c$. The boundary conditions are no--flux for $p$ and a fixed potential difference of 20 ($V=20$) for $\phi$. Both cases demonstrate in addition, crowding and over--screening effects near the boundaries.}\label{fig:RTILmodel}
\end{figure}
\begin{figure*}
	\includegraphics[width=5.5in]{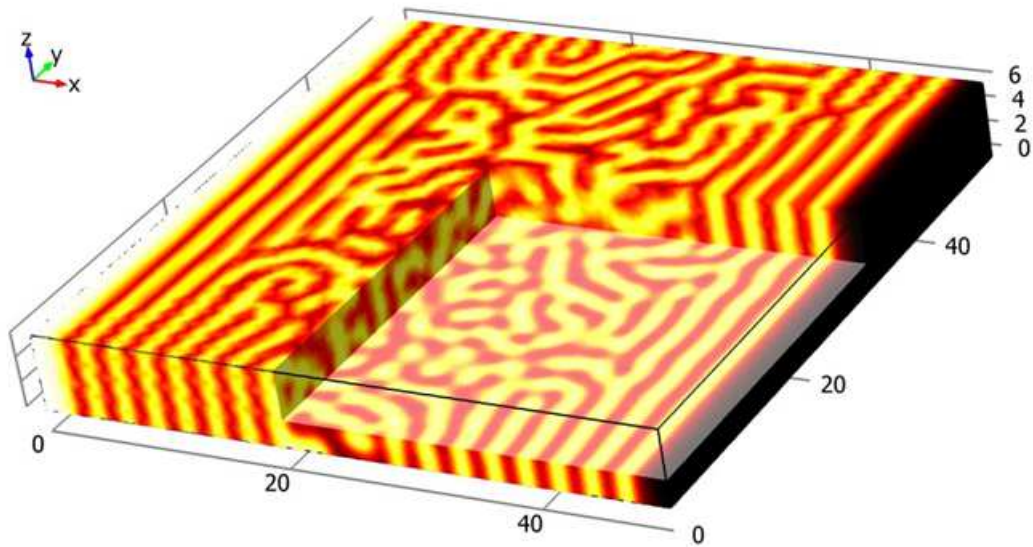}
	\caption{Numerical solution of~\eqref{eq:rtil_model} in 3D dimensions, showing $p(x,y,z)$ for $\sigma=0.24<\sigma_c(\chi=3)= 1/4$, and potential along the $x$ axis with $V=4$. Light and dark colors mark the upper ($p=1$) and the lower ($p=0$) limits, respectively.}\label{fig:RTILmodel3D}
\end{figure*}

Next, we consider a finite domain and apply boundary conditions \eqref{eq:BC}. Indeed, numerical integrations of~\eqref{eq:rtil_model_p_1D} show that, in addition to the EDL crowding and over--screening effects (cf.~\cite{Yochelis2014b}), there is a bifurcation to a periodic structure, as demonstrated in Figure~\ref{fig:RTILmodel}. The bifurcation is of a supercritical type, i.e., the amplitude of the emerging solutions scales as $p \sim \sqrt{\sigma_c-\sigma}$ (details will be given elsewhere), and emerge also \textit{without} any potential difference at the walls~\cite{Cross1993851}. Notably, finite wavenumber bifurcation cannot arise in any extension of Poisson-Nernst-Planck framework~\cite{Bazant2011,Yochelis2014b,Lee2015}, namely, the bulk will preserve spatial symmetry also in higher space dimensions. 

On the other hand, in~\eqref{eq:rtil_model}, the finite wavenumber instability persists on large 2D and 3D domains ($d \gg \lambda$) but gives rise to a labyrinthine pattern rather than stripes, as shown in Fig.~\ref{fig:RTILmodel3D}. We note though that, integration of~$p(x,y,z)$ over $(y,z)$ will result in a vanishing charge density, so that the resulting $\bar{p}(x)$ profile will resemble the 1D stationary solution below the instability onset, e.g., Fig.~\ref{fig:RTILmodel}a. However, even though the asymptotic solutions may resemble the 1D case and also the profiles obtained via FPNP theory, it is a distinct solution with distinct properties, for example the respective time scales upon approaching them will differ. Recovering these distinct time scales in~\eqref{eq:rtil_model} is a prerequisite to a systematic comparison with empirical observations and for determining strategies to tailor by-demand RTIL compositions.

{\em Discussion--}
Electrical diffuse layer in RTILs displays a puzzling charge layering near the liquid/solid interface, which vanishes toward the bulk region. This unique property stimulated recently intensive studies and distinct approaches ranging from electrode polarization effects to bulk morphology based descriptions~\cite{fedorov2014ionic,hayes2015structure}. However, since understanding of EDL properties is required to control and optimize charge transport and transfer in numerous energy conversion devices, a study of the mechanisms governing RTIL structure was moved to a spotlight~\cite{Yochelis2015}. To capture both the polarization effects and the bulk properties, we consider a mean-field framework which couples short-range interactions (e.g., finite-size effects) and long-range Coulombic interactions between the ions. Specifically, we use phase separation of Cahn-Hilliard type coupled to Poisson equation
to unfold the origin of the emerged spatially periodic and isotropic bulk morphology, which becomes then ordered (anisotropic) near the electrode surface (see Fig.~\ref{fig:RTILmodel3D}), cf.~\cite{Yochelis2015,hayes2015structure}.

Contrary to the currently facilitated Fermi-Poisson-Nernst-Planck approach~\cite{Bazant2011,Lee2015}, which is based on the concentrated electrolyte description ($p(\mathbf{x})+n(\mathbf{x}) < 1$), our Cahn-Hilliard-Poisson approach keeps a qualitative fidelity to the physicochemical empirical observations, within an Onsager system framework. In concentrated electrolytes most of the volume is occupied by water and thus, the mass transport is attributed to ion diffusion with Bikerman's modification to capture finite size effects~\cite{Bikerman1942,Borukhov1997435,Bazant2011}. Instead, we account for finite-size effects using the Cahn-Hilliard energy, i.e., via phase separation. In this case, the ion transport is governed by the inter-diffusion while Coulombic interactions enter naturally through the Poisson equation, unlike the short-range electrostatic correction used in the Fermi-Poisson-Nernst-Planck approach~\cite{Bazant2011,Lee2015}. Surprisingly, although Coulombic interactions are long-ranged, their coupling to the phase separation results in a finite wavenumber instability that is a characteristic of activator--inhibitor Turing type systems. Notably, this behavior cannot arise in variants of Poisson-Nernst-Planck framework. Nevertheless, RTILs can be diluted by either organic solvent or by formation of neutral ion-pairs~\cite{Yochelis2015}, and so the connection to a ternary system that comprises in addition, a solvent is a natural extension that will be presented elsewhere.

In general, our formulation bears similarity to the well-studied Ohta-Kawasaki energy  for diblock copolymer mixtures and implies that RTILs and diblock copolymers share common morphologies, such as lamellar, spheres, circular tubes, and bicontinuous gyroids patterns, as has been already conjectured in~\cite{Yochelis2015}. In particular, since RTIL ions are not chemically bonded like the diblock copolymers, the interplay between morphology and long-range interactions (electrokinetics in the case of RTILs) is richer and will be studied in detail elsewhere.  As such, we expect that the platform developed here, can be extend to a much wider range of material science and energy conversion applications that rely on the optimized coupling between material nanostructure, electrostatics, electrodiffusion, and nanoflows.

{\em Acknowledgements--} This research was done in the framework of the Grand Technion Energy Program (GTEP) and of the BGU Energy Initiative Program, and supported by the Adelis Foundation for renewable energy research.  N.G. acknowledges the support from the Technion VPR fund and from EU Marie--Curie CIG grant 2018620.

\bibliography{PNP}
 \end{document}